\documentclass[%
 aip,
 amsmath,amssymb,
 reprint,%
]{revtex4-1}

\usepackage{graphicx}
\usepackage{dcolumn}
\usepackage{bm}
\usepackage[utf8]{inputenc}
\usepackage[T1]{fontenc}
\usepackage{mathptmx}
\usepackage{etoolbox}
\usepackage{caption}
\usepackage{subcaption}
\usepackage{url}            
\usepackage{subfiles}
\usepackage{array}
\newcolumntype{P}[1]{>{\centering\arraybackslash}m{#1}}
\usepackage{lineno}
\usepackage{microtype}
\usepackage{amsmath}
\usepackage{amssymb}
\usepackage{mathtools}
\makeatletter
\def\@email#1#2{%
 \endgroup
 \patchcmd{\titleblock@produce}
  {\frontmatter@RRAPformat}
  {\frontmatter@RRAPformat{\produce@RRAP{*#1\href{mailto:#2}{#2}}}\frontmatter@RRAPformat}
  {}{}
}%
\makeatother
\begin{document}

\preprint{AIP/123-QED}

\title{Accurate simulation of direct laser acceleration in a laser wakefield accelerator}
\author{Kyle G. Miller}
\email{kmill@lle.rochester.edu}
\affiliation{Laboratory for Laser Energetics, University of Rochester, Rochester, NY 14623-1299, USA}

\author{John P. Palastro}
\affiliation{Laboratory for Laser Energetics, University of Rochester, Rochester, NY 14623-1299, USA}

\author{Jessica L. Shaw}
\affiliation{Laboratory for Laser Energetics, University of Rochester, Rochester, NY 14623-1299, USA}

\author{Fei Li}
\affiliation{Department of Physics and Astronomy, University of California, Los Angeles, CA 90095, USA}

\author{Frank S. Tsung}
\affiliation{Department of Physics and Astronomy, University of California, Los Angeles, CA 90095, USA}

\author{Viktor K. Decyk}
\affiliation{Department of Physics and Astronomy, University of California, Los Angeles, CA 90095, USA}

\author{Warren B. Mori}
\affiliation{Department of Physics and Astronomy, University of California, Los Angeles, CA 90095, USA}

\date{\today}

\begin{abstract}
In a laser wakefield accelerator (LWFA), an intense laser pulse  excites a plasma wave that traps and accelerates electrons to relativistic energies. When the pulse overlaps the accelerated electrons, it can enhance the energy gain through direct laser acceleration (DLA) by resonantly driving the betatron oscillations of the electrons in the plasma wave. The particle-in-cell (PIC) algorithm, although often the tool of choice to study DLA, contains inherent errors due to numerical dispersion and the time staggering of the electric and magnetic fields.  Further, conventional PIC implementations cannot reliably disentangle the fields of the plasma wave and laser pulse, which obscures interpretation of the dominant acceleration mechanism. Here, a customized field solver that reduces errors from both numerical dispersion and time staggering is used in conjunction with a field decomposition into azimuthal modes to perform PIC simulations of DLA in an LWFA. Comparisons with traditional PIC methods, model equations, and experimental data show improved accuracy with the customized solver and convergence with an order-of-magnitude fewer cells. The azimuthal-mode decomposition reveals that the most energetic electrons receive comparable energy from DLA and LWFA.
\end{abstract}


\maketitle

\section{Introduction} \label{sec:intro}
Laser wakefield accelerators (LWFAs) use intense laser pulses to excite plasma waves that can trap and accelerate electrons to relativistic energies over short distances.\cite{Tajima1979} While the acceleration is predominantly in the longitudinal direction, the trapped electrons also undergo transverse, i.e., betatron, oscillations in the fields of the plasma wave. In an idealized LWFA, an ultrashort laser pulse drives the plasma wave but does not overlap or interact with the trailing trapped electrons.\cite{Lu2007,Esarey2009} There are, however, a number of situations in which this simple picture breaks down. For instance, the accelerated electrons can travel faster than the group velocity of the laser pulse and move forward in the plasma wave into the laser light. The laser light can also slide backward into the accelerated electrons when the pulse depletes and locally decelerates. Experimentally, the laser bandwidth places a lower bound on the duration of the pulse that can be comparable to or longer than the plasma period. As an example, a laser pulse with a duration longer than the plasma wave period can drive a self-modulated LWFA (SM-LWFA). Here, the pulse inherently overlaps the accelerated electrons and can even create an ion channel where electrons are accelerated purely by the fields of the laser pulse.\cite{Joshi1981,Krall1993,Modena1995,Pukhov2002,Pukhov2003,Mangles2005} Finally, proposed ionization injection and enhanced acceleration schemes for LWFAs employ two laser pulses, with the trailing pulse colocated with the trapped electrons.\cite{Xu2014LowWavelengths,Zhang2015}

In each of these examples,  the electric field of the laser pulse can resonantly drive the  betatron oscillations of the accelerated electrons. Specifically, betatron resonance occurs when the Doppler-shifted laser frequency observed by an accelerated electron matches its betatron frequency, $\omega_\beta$, i.e.,
\begin{equation} \label{eq:matching}
(1-\beta_z)\omega_0 = \frac{\omega_p}{\sqrt{2\gamma}} \equiv \omega_\beta, 
\end{equation}
where $\omega_0$ is the laser frequency, $\beta_z$ is the longitudinal electron velocity, $\omega_p$ is the plasma frequency, and $\gamma$ is the relativistic factor of the electron. When Eq.~(\ref{eq:matching}) is (approximately) satisfied, electrons can gain energy directly from the laser pulse in a process known as direct laser acceleration (DLA).\cite{Pukhov1999,Gahn1999,Mangles2005,Arefiev2015,Khudik2016} The transverse momentum gained from the  electric field of the pulse is rotated into longitudinal momentum by the magnetic field of the pulse.

Particle-in-cell (PIC) simulations of DLA can guide experiments and assist in interpretation of the results, but accurate simulations remain a challenge. The phasespace and energy gain of electrons in betatron resonance are especially sensitive to the phase and group velocities of the laser pulse as well as the relative amplitudes of the electric and magnetic fields. This presents an issue for traditional PIC methods, which exhibit numerical dispersion errors for light waves and field-amplitude errors due to the time staggering of the electric and magnetic fields.\cite{Yee1966,Li2021,Tangtartharakul2021ParticleInteractions} The time-staggering error can be mitigated by subcycling and temporal interpolation of the fields, but dispersion errors would persist.\cite{Tangtartharakul2021ParticleInteractions} Several spectral methods feature a more-precise dispersion relation for light waves,\cite{Dawson1983,Birdsall1984,Liu1997TheWavelength} but these methods are typically less amenable to massive parallelization. An exception is the pseudo-spectral time-domain algorithm presented by Vay \emph{et al}.\cite{Vay2013} To address the need for accuracy when using finite-difference solvers, a customized explicit Maxwell field solver was recently developed\cite{Li2021} that corrects for errors in the dispersion relation of light waves and the time-staggering error in the Lorentz force.  Simulations employing this customized solver have produced electron motion in the fields of a laser pulse that more-closely align with exact theoretical solutions than simulations using the standard Yee solver.\cite{Li2021}

Interpreting and optimizing the energy gain of electrons in DLA requires disentangling the contributions from the laser pulse and the plasma wave. These contributions are often calculated by attributing the work done by the longitudinal and perpendicular electric fields to the plasma wave and laser pulse, respectively.\cite{Gahn1999,Pukhov2003,Zhang2016,Shaw2016EstimationSimulations,Shaw2017,Shaw2018}  However, focused laser pulses can have a non-negligible longitudinal electric field that has been shown to reduce the energy gain of electrons in betatron resonance.\cite{Pukhov2003,King2021,Wang2019} Alternatively, one can make use of a cylindrical domain decomposition into azimuthal modes,\cite{Lifschitz2009,Davidson2015,Lehe2016,Wang2019} which naturally splits the cylindrically symmetric wakefields and fields of a linearly polarized laser pulse into separate quantities.  This so-called ``quasi-3D'' decomposition enables accurate calculations of the work performed by the fields of the laser pulse and plasma wave, including any contribution from the longitudinal field of the pulse.  In addition, the quasi-3D geometry provides 3D geometric effects at the computational cost of a 2D simulation, expediting optimization and design scans in support of experiments.

In this work, we employ the recently developed customized solver\cite{Li2021} and a cylindrical decomposition into azimuthal modes\cite{Davidson2015} to study electron motion and DLA in an LWFA. Comparisons with a reduced physical model and experimental results show much better agreement when using the customized solver compared to standard methods. Convergence tests show that the customized solver produces reliable results with an order-of-magnitude larger cell size than the standard Yee solver. The azimuthal-mode expansion allows for a precise partitioning of the work done on electrons into contributions from the laser pulse and plasma wave. Specifically, two populations of accelerated electrons are observed, one receiving energy purely from the wakefields and another attaining higher energies from the combined effects of wakefield and direct laser acceleration. More generally, the combined computational savings from the customized solver and the quasi-3D decomposition enables simulation of larger and more-complex physical problems and clearer analysis of acceleration mechanisms in laser--plasma interactions.

\section{Physical model for direct laser acceleration} \label{sec:betatron}
Electrons trapped in the plasma wave of an LWFA can be accelerated by the fields of both the plasma wave and the laser pulse. The wakefields of the plasma wave accelerate the electrons longitudinally, while also providing a transverse focusing force that causes the electrons to undergo betatron oscillations. When the laser pulse overlaps these electrons and the betatron resonance condition in Eq.~\eqref{eq:matching} is satisfied, the transverse electric field of the laser pulse is in phase with the electron betatron motion and can increase the transverse momentum.  
The increased transverse momentum is then rotated into longitudinal momentum by the magnetic field of the pulse, resulting in a steady energy gain from the pulse that augments the energy gained from the longitudinal wakefield.\cite{Pukhov2002,Shaw2017,Wang2019} This section presents a model for the electron motion in the combined fields of the plasma wave and laser pulse, which will be used to assess the impact of errors inherent to certain PIC algorithms.

Consider a laser pulse polarized in the $\bold{\hat{x}}$~direction and propagating in the $\bold{\hat{z}}$~direction with longitudinal wavenumber $k_z$ and phase velocity $\mathrm{v}_\phi=\omega_0/k_z$. The pulse travels at its group velocity and exerts a ponderomotive force that expels electrons from its path, creating a bubble of net positive charge surrounded by a thin electron sheath. The bubble travels at a velocity $\mathrm{v}_b \lesssim c$.

The vector potential of the laser pulse  $\bold{A}_l$ will be modeled using the Coulomb gauge, i.e, $\nabla \cdot \bold{A}_l = 0$. The transverse vector potential of the laser pulse $A_{x,l}$ satisfies the Helmholtz equation inside the bubble, which acts like a cylindrical waveguide for the pulse. Specifically, \begin{equation} \label{eq:laserchannelAx}
		A_{x,l} = A_0 J_0(k_\perp r)\cos(k_z\xi),
\end{equation}
where $J_n$ denotes the $n^\mathrm{th}$ cylindrical Bessel function, $r=(x^2+y^2)^{1/2}$, $\xi \equiv z-\mathrm{v}_\phi t$, and $k_z = (\omega_0^2/c^2 - k_\perp^2)^{1/2}$. A value of $k_\perp = 2.4/(\bar{r}_b +c/\omega_p)$ ensures that the Bessel function vanishes at the edge of the electron sheath outside the bubble and $\bar{r}_b$ denotes the average bubble radius in the region of the accelerated electron bunch. With Eq.~\eqref{eq:laserchannelAx}, the longitudinal vector potential of the laser pulse can be calculated exactly using $A_{z,l} = \int \partial_x A_{x,l} \mathrm{d}z$: \begin{equation} \label{eq:laserchannelAz}
		A_{z,l} = A_0\frac{k_\perp x}{k_zr} J_1(k_\perp r)\sin(k_z\xi),
\end{equation}
where the integration constant is chosen so that the potential vanishes as $r \rightarrow \infty$. 

The vector $\bold{A}_b$ and scalar $\phi_b$ potentials associated with the bubble 
are more-naturally modeled in 
the Lorenz gauge, i.e., $c^2\nabla \cdot \bold{A}_b +\partial_t \phi_b = 0$. Following the analysis of Lu \emph{et al}.,\cite{Lu2006} the potentials within the bubble are given by 
\begin{equation} \label{eq:wake-potentials}
\begin{aligned}
    &\psi = \frac{\omega_p^2 m}{4e}\left[\big(1+\beta(r_b)\big)r_b^2-r^2  \right], \\
    &\phi_b = \phi_0(\zeta)-\frac{\omega_p^2 m}{4e}r^2, \\
     &A_{z,b}=A_{z,0}(\zeta), \\
     &A_{r,b}=\frac{r}{2c} \frac{d\psi_0}{d\zeta},
\end{aligned}
\end{equation}
where $\psi\equiv \phi_b-c A_{z,b}$ is the wake potential, $m$ is the electron mass, $e$ the elementary charge, $r_b(\zeta)$ the blowout or bubble radius as a function of $\zeta = z - ct$, $\beta(r_b)$ a function that depends on the details of the electron sheath that surrounds the ion column,\cite{Lu2006,Lu2006NonlinearRegime} and for any quantity $S$, $S_0(\zeta) \equiv S(\zeta,r=0)$. The trajectory $r_b(\zeta)$ is determined by a nonlinear differential equation [see Eq.~(46) and subsequent paragraph in Ref.~\onlinecite{Lu2006}]. In the limit that $\omega_p r_b/c \gg 1$, this equation predicts that the bubble has a near-spherical shape, \cite{Lu2006,Lu2006NonlinearRegime} and the resulting expression for $\psi$ approaches that described by Kostyukov \emph{et al}.:\cite{Kostyukov2004PhenomenologicalRegime} 
\begin{equation}
\label{eq:bubblepotential}
\psi \approx \frac{\omega_p^2 m}{4e}(r_{b,\mathrm{max}}^2-\zeta^2-r^2),
\end{equation}
where $r_{b,\mathrm{max}}$ is the maximum blowout radius and $\zeta$ is defined such that $r_b(\zeta=0)=r_{b,\mathrm{max}}$. A gauge transformation can then be made such that $A_{z,b} = -\phi_b/c$ and $A_{\perp,b}=0$.
 
 With the potentials known, the electric and magnetic fields can be calculated in the usual way: $\bold{E} = -\partial_t (\bold{A}_l + \bold{A}_b) - \nabla \phi_b$ and $\bold{B} = \nabla \times (\bold{A}_l + \bold{A}_b)$. The momenta of the accelerated electrons evolve in response to the resulting Lorentz force, 
\begin{equation} \label{eq:motion1}
	\frac{d\bold{p}}{dt} = -e (\bold{E} + \bold{v} \times \bold{B}),
\end{equation}
where $\bold{p} = \gamma m \bold{v}$ and $\bold{v}$ is the electron velocity. Equation~\eqref{eq:motion1} can be re-expressed as equations of motion for the coordinates of an electron:
\begin{equation} \label{eq:motionchannel}
\begin{aligned}
\ddot{x} + \frac{\dot{\gamma}}{\gamma}\dot{x} + \omega_\beta^2 \left[ 1 + \frac{e}{\omega_p^2 m}\frac{c-\dot{z}}{c}\frac{d^2\psi_0}{d\zeta^2} \right] x &= \frac{F_{x,l}}{\gamma m}, \\
\ddot{y} + \frac{\dot{\gamma}}{\gamma}\dot{y} + \omega_\beta^2 \left[ 1 + \frac{e}{\omega_p^2 m}\frac{c-\dot{z}}{c}\frac{d^2\psi_0}{d\zeta^2} \right] y &= \frac{F_{y,l}}{\gamma m}, \\
\ddot{\zeta} + \frac{\dot{\gamma}}{\gamma}(\dot{\zeta}+c) - \frac{\omega_\beta^2e}{\omega_p^2 m} \left[2\frac{d\psi_0}{d\zeta}-\frac{x\dot{x}+y\dot{y}}{c}\frac{d^2\psi_0}{d\zeta^2} \right] &= \frac{F_{z,l}}{\gamma m},
\end{aligned}
\end{equation}
where a dot denotes a full time derivative, $\bold{F}_l = -e\left( \bold{E}_l + \bold{v} \times \bold{B}_l \right)$, and $\bold{E}_l$ and $\bold{B}_l$ are the electric and magnetic fields of the laser pulse. These equations are general and can describe motion in bubbles or in pure ion channels with the appropriate expression for $\psi_0$. Upon taking the limit of a spherical bubble, using Eqs.~(\ref{eq:bubblepotential}), (\ref{eq:laserchannelAx}), and (\ref{eq:laserchannelAz}), and restricting motion to the $x$--$z$ plane (i.e., $y=0$), the system in Eq.~(\ref{eq:motionchannel}) reduces to
\begin{widetext}
\begin{equation} \label{eq:motionchannelreduced}
\begin{aligned}
\ddot{x} + \frac{\dot{\gamma}}{\gamma}\dot{x} + \frac{\omega_\beta^2}{2} \left[ 1 + \frac{\dot{z}}{c} \right] x  &=
  \frac{a_0 c}{\gamma} \Bigg[ (\omega_0 -  k_z \dot{z}) J_0(k_\perp x) - \frac{k_\perp^2}{k_z^2} k_z \dot{z}  J_1'(k_\perp x) \Bigg] \sin(k_z \xi), \\
\ddot{\zeta} + \frac{\dot{\gamma}}{\gamma}(\dot{\zeta}+c) + \omega_\beta^2 \left[\zeta-\frac{x\dot{x}}{2c} \right] &=
  - \frac{a_0 c}{\gamma} \Bigg\{ \frac{k_\perp}{k_z} \omega_0 J_1(k_\perp x) \cos(k_z \xi) -  k_z \dot{x} \left[ J_0(k_\perp x) + \frac{k_\perp^2}{k_z^2} J_1'(k_\perp x) \right] \sin(k_z\xi) \Bigg\},
\end{aligned}
\end{equation}
\end{widetext}
where $J_1'(s) \equiv d J_1(s)/d s$ and $a_0 = \frac{e A_0}{m c}$ is the normalized vector potential of the laser pulse. For a cylindrically symmetric laser pulse, an electron initialized at $y=0$ will experience no force in the $\hat{\bold{y}}$ direction, justifying the consideration of motion in only the $x$--$z$ plane.

In Eqs.~(\ref{eq:wake-potentials})--(\ref{eq:motionchannelreduced}), the bubble velocity $\mathrm{v}_b$ has been approximated as $c$. More generally, the wake potentials have a weak dependence on $\mathrm{v}_b$. Single-particle calculations that maintained this dependence exhibited negligible differences from cases where $\mathrm{v}_b$ was set to the vacuum speed of light. The results are, however, sensitive to the value of $\mathrm{v}_{\phi}$. Thus, the more important effect is the phase slippage of electrons within the fields of the laser pulse.

Nemeth \emph{et al}.\cite{Nemeth2008} numerically solved different equations of motion for electrons in the combined fields of a bubble and a planewave laser pulse, comparing the results to a PIC simulation. The phase velocity used in the equations of motion was extracted from the PIC simulation. The calculations and PIC results agreed; however, the phase velocity was artificially low due to dispersion errors in the PIC algorithm used. In contrast, the study described here numerically solves Eq.~\eqref{eq:motionchannelreduced} with a physically accurate phase velocity extracted from a PIC simulation that uses a customized, dispersion-free finite-difference field solver that conserves charge. This solver is described in the next section (Sec.~\ref{sec:numerics}), and comparisons of the numerically integrated electron trajectories with those from the PIC simulation are presented in Sec.~\ref{sec:pic-integrated}.

\section{Numerical methods for the particle-in-cell algorithm} \label{sec:numerics}
Despite the utility and widespread adoption of the PIC algorithm for studying laser--plasma interactions such as LWFA, many of the standard numerical methods come with significant errors or can complicate analysis. In this section, recent techniques are described that can more-accurately model and simplify the analysis of electron dynamics in a laser pulse, such as DLA-assisted LWFA.

\subsection{Customized ``dual'' solver} \label{sec:solver}

The use of the finite-difference time-domain technique for updating the electromagnetic fields makes the PIC method susceptible to numerical issues such as improper numerical dispersion,\cite{Lehe2013NumericalAcceleration} numerical Cerenkov radiation,\cite{Lehe2013NumericalAcceleration,Godfrey2013NumericalAlgorithm,Xu2013} and finite-grid instability.\cite{Langdon1970EffectsPlasmas,Meyers2015OnInstability}  These errors often depend on the resolution but are not always eliminated when the time step and/or cell sizes are decreased.  A customized, higher-order solver was recently developed\cite{Li2021} to eliminate the dual errors of improper numerical dispersion of light in vacuum and the incorrect Lorentz force on particles within a laser field due to the time staggering of electric and magnetic fields. The full details of the solver are described in Ref.~\onlinecite{Li2021}, but the algorithm is summarized here.

The electromagnetic field update in a typical PIC algorithm is performed by solving Faraday's and Ampere's laws.  Using the standard Yee discretization,\cite{Yee1966} the electric and magnetic fields, $\bold{E}$ and $\bold{B}$, respectively, are staggered in space and time to allow for central finite-difference approximations to the first-order derivatives that are accurate to second order in space and time. However, the finite-differences in space can be altered to obtain more-accurate solutions, and separate, unique finite-difference operators can be applied to $\bold{E}$ and $\bold{B}$. Taking the Fourier transform of Faraday's and Ampere's laws in space and time yields
\begin{equation} \label{eq:Maxwell}
    \begin{gathered}
        [\omega]_t \tilde{\bold{B}} = [\bold{k}]_E \times \tilde{\bold{E}}, \\
        [\omega]_t \tilde{\bold{E}} = -c^2[\bold{k}]_B \times \tilde{\bold{B}},
    \end{gathered}
\end{equation}
where $[\cdot]_*$ represents the Fourier transform of finite-difference derivatives with respect to time $t$ or with respect to space on $\bold{E}$ and $\bold{B}$, respectively. These equations can be combined with Gauss's law, $i[\bold{k}]_B \cdot \tilde{\bold{E}} = 0$, to provide
\begin{equation} \label{eq:disp_tot}
    [\omega]_t^2 - c^2[\bold{k}]_B \cdot [\bold{k}]_E = 0,
\end{equation}
which is the numerical dispersion relation for a given finite-difference scheme.  For the central finite-difference operator in time, $[\omega]_t = \sin \left( \frac{\omega \Delta t}{2} \right) / \frac{\Delta t}{2}$, where $\Delta t$ is the time step. For a light wave traveling along the $\hat{\bold{z}}$~direction with wavenumber $k_z$, the $[k_z]_{B}$ and $[k_z]_{E}$ operators are to be constructed such that $\omega = ck_z$.  With this requirement, Eq.~(\ref{eq:disp_tot}) can be rewritten as
\begin{equation} \label{eq:cond1}
    [k_z]_{B}[k_z]_{E} = \left[ \sin \left( \frac{ck_z \Delta t}{2} \right) \middle/ \frac{c\Delta t}{2} \right]^2,
\end{equation}
which is the first condition on the finite-difference operators $[k_z]_{B}$ and $[k_z]_{E}$.

The second condition eliminates errors in the Lorentz force of a laser pulse due to the time staggering of the $\bold{E}$ and $\bold{B}$ fields.  The electromagnetic force computed on a particle is time-centered with $\bold{E}$ and the particle position, meaning that the transverse Lorentz force from an electromagnetic wave polarized in the $\hat{\bold{x}}$~direction is computed at time step $n$ as
\begin{equation} \label{eq:Lorentz-real}
    F_x^n = q \left( E_x^n - c\bar{\beta}_z^n \bar{B}_y^n \right),
\end{equation}
where the superscript $n$ denotes a quantity at time $n\Delta t$, $q$ is the particle charge, $\beta_z = \mathrm{v}_z/c$, and the overbar indicates an average in time, i.e., $\bar{\left< \cdot \right>}^n = \frac{1}{2} (\left< \cdot \right>^{n-\frac{1}{2}} + \left< \cdot \right>^{n+\frac{1}{2}})$.  The corresponding Fourier transform is 
\begin{equation} \label{eq:Lorentz-fouier}
    \tilde{F}_x = q \left( \tilde{E}_x - c\bar{\beta}_z^n \tilde{B}_y \cos \frac{\omega \Delta t}{2} \right) \tilde{S},
\end{equation}
where $\sim$ indicates a quantity in the $(\omega, \bold{k})$ domain and $S$ is the interpolation function.\cite{Li2021}  The magnetic field in Eq.~(\ref{eq:Lorentz-fouier}) can be eliminated by making use of Faraday's law in Eq.~(\ref{eq:Maxwell}) along with Eq.~(\ref{eq:disp_tot}) to give
\begin{equation} \label{eq:Lorentz-ke-kb}
    \tilde{F}_x = q \tilde{E}_x \left( 1 - \bar{\beta}_z^n \sqrt{\frac{[k_z]_{E}}{[k_z]_{B}}} \cos \frac{\omega \Delta t}{2} \right) \tilde{S}.
\end{equation}
Thus, the correct Lorentz force on a particle (assuming $\omega = c k_z$) requires
\begin{equation} \label{eq:cond2}
    \sqrt{\frac{[k_z]_{E}}{[k_z]_{B}}} \cos \left(\frac{c k_z \Delta t}{2} \right) = 1,
\end{equation}
which is the second condition on the finite-difference operators $[k_z]_{B}$ and $[k_z]_{E}$.

A customized solver can be created with finite-difference coefficients such that it solves Eqs.~(\ref{eq:cond1}) and (\ref{eq:cond2}) to desired precision, as was done in Ref.~\onlinecite{Li2021}.  This solver is referred to as the ``dual'' solver because it simultaneously addresses the issues of inaccurate numerical dispersion and time-staggering errors in the Lorentz force. An important feature of the dual solver is that only the temporal resolution appears in Eqs.~(\ref{eq:cond1}) and (\ref{eq:cond2}), as opposed to both the spatial and temporal resolutions as is the case with the Yee solver.\cite{Lehe2013NumericalAcceleration,Li2021}  This allows the time step to be reduced independently of the spatial discretization to obtain greater accuracy. In contrast, dispersion errors become more severe when using the Yee solver for time steps smaller than the Courant--Friedrichs--Lewy (CFL) condition.

\subsection{Azimuthal-mode expansion} \label{sec:quasi-3D}

Analyzing the acceleration and energy gain of an electron in laser--plasma interactions requires differentiating the work done by the laser pulse from the work done by the plasma waves. In studies of DLA-assisted LWFA, this distinction is often made by attributing the work done by the laser pulse and plasma wave to that done by the perpendicular and longitudinal fields, respectively.\cite{Gahn1999,Pukhov2003,Zhang2016,Shaw2017} This approach, however, misattributes work done by the longitudinal field of the laser pulse to the plasma wave.
Azimuthal decomposition of the fields provides a more-natural and physically accurate way to separate the contributions from the laser pulse and plasma wave.  

In the azimuthal-mode decomposition, or ``quasi-3D'' geometry,\cite{Davidson2015} fields are expanded as a series in the azimuthal mode number $m$ with amplitudes that vary as $e^{im\phi}$, where $\phi$ is the azimuthal angle. As a result, the axisymmetric $m=0$ fields of the plasma wave (both transverse and longitudinal) are completely distinguishable from the $m=1$ fields of the laser pulse. The work done by the plasma wave and laser pulse can then be calculated as
\begin{equation} \label{eq:work-modal}
	\begin{gathered}
		W_\mathrm{LWFA} = -e \int \bold{E}^0 \cdot \bold{v} \, dt, \\
		W_\mathrm{DLA} = -e \int \bold{E}^1 \cdot \bold{v} \, dt,
	\end{gathered}
\end{equation}
where the superscript denotes the azimuthal mode. Note that a similar separation was previously applied to the results of simulations based on the quasistatic approximation,\cite{Wang2019} but only recently has this technique been applied to fully self-consistent PIC models,\cite{King2021} which are required to model processes such as self injection.

While the calculation in Eq.~(\ref{eq:work-modal}) is exact for the quasi-3D algorithm, estimates can be made in Cartesian geometries to improve the accuracy of a separation in terms of the field components. The work done by the longitudinal and transverse fields is given by
\begin{equation} \label{eq:work-trad}
	\begin{gathered}
		W_z = -e \int E_z \mathrm{v}_z \, dt = -e \sum_m \int E_z^m \mathrm{v}_z \, dt, \\
		W_\perp = -e \int \bold{E}_\perp \cdot \bold{v}_\perp \, dt = -e \sum_m \int \bold{E}_\perp^m \cdot \bold{v}_\perp \, dt.
	\end{gathered}
\end{equation}
The transverse electric fields of the plasma wave do very little net work, but the longitudinal electric field of the laser pulse can contribute substantially to $W_z$.  Near focus, the field components of a linearly polarized laser pulse with a transverse Gaussian profile can be approximated as\cite{Quesnel1998}
\begin{equation} \label{eq:laser-fields}
    \begin{gathered}
        \bold{\hat{x}} \cdot \bold{E}_L = E_0 \exp\left(-\frac{r^2}{w_0^2}\right) \sin \left( \theta \right), \\
        \bold{\hat{z}} \cdot \bold{E}_L = \frac{2 x}{k_z w_0^2} E_0 \exp\left(-\frac{r^2}{w_0^2}\right) \cos \left( \theta \right),
    \end{gathered}
\end{equation}
where $E_0$ is the peak electric-field amplitude, $w_0$ is the spot size, and $\theta$ is the phase as seen by an electron.  Suppose the electron is moving forward with a velocity $\mathrm{v}_z \approx c$ and executing small-amplitude transverse betatron oscillations, i.e., $x(t) = A \cos{\omega_\beta t}$ with $|A| \ll w_0$.  If the electron is in betatron resonance, the Doppler-shifted laser frequency matches the betatron frequency: $\theta = \omega_\beta t$.  The cycle-averaged work done by the laser pulse due to its longitudinal and transverse components can then be calculated from 
\begin{equation}
    \begin{gathered}
        \langle W_z \rangle = -e \int_0^\frac{2\pi}{\omega_\beta} (\bold{\hat{z}} \cdot \bold{E}_L)\mathrm{v}_z\,dt, \\
        \langle W_\perp \rangle = -e \int_0^\frac{2\pi}{\omega_\beta} (\bold{\hat{x}} \cdot \bold{E}_L)\mathrm{v}_x\,dt.
    \end{gathered}
\end{equation}
Taking the ratio of these two quantities yields\cite{Pukhov2003}
\begin{equation} \label{eq:f}
    f \equiv \frac{\left\langle W_z \right\rangle}{\left\langle W_\perp \right\rangle} = -\frac{2c^2\sqrt{2\gamma}}{\omega_0 \omega_p w_0^2}.
\end{equation}
Note that the ratio $f$ is negative, indicating that when an electron is in phase with the transverse component of a linearly polarized Gaussian laser pulse, it is out of phase with the longitudinal component.

If one does not have access to an azimuthal decomposition of the fields, Eq.~(\ref{eq:f}) can be used to estimate the work done by the LWFA and DLA processes:
\begin{equation} \label{eq:work-f}
    \begin{gathered}
        W_\mathrm{LWFA}^* = W_z - fW_\perp, \\
        W_\mathrm{DLA}^* = (1+f)W_\perp.
    \end{gathered}
\end{equation}
This estimate was previously compared to Eq.~(\ref{eq:work-modal}) using results from an SM-LWFA simulation in the quasi-3D geometry, and the agreement was reasonable.\cite{King2021}  However, the accuracy of Eq.~(\ref{eq:f}) relies on the assumptions that $\mathrm{v}_z \approx c$ and that the oscillation amplitude is much less than the spot size. More generally, electrons can have large transverse velocities, and the amplitude of oscillation can extend to the edge of the bubble. In addition, the spot size can vary greatly with longitudinal position, as can the local frequency due to photon acceleration.\cite{Mori1997,Zhu2013PulsedSimulations}  The accuracy of the approximation in Eq.~(\ref{eq:work-f}) is compared with Eq.~(\ref{eq:work-modal}) in Sec.~\ref{sec:pic-energy}.

\section{Simulation results and discussion} \label{sec:results}

This section presents full-scale 3D and quasi-3D simulations of an LWFA where the laser pulse fills the first bubble. The simulations are motivated by the experiments described in Refs.~\onlinecite{Shaw2017,Shaw2018}. In the experiments, the plasma density was varied to generate electron spectra with and without signatures of DLA, and tunneling ionization of nitrogen-doped helium\cite{Pak2010,Oz2007,McGuffey2010IonizationAccelerator} was used to inject electrons into the bubble.
In one case from Ref.~\onlinecite{Shaw2017}, DLA was observed to contribute substantially to the electron energy gain.  The simulations presented here model this case using a laser pulse polarized in the $\bold{\hat{x}}$~direction that propagates in the positive $\bold{\hat{z}}$~direction through a mixture of 99.9\%~He and 0.1\%~N$_2$ (see Table~\ref{tab:params} and Appendix~\ref{app:params} for the physical and numerical parameters, respectively). The vast majority of trapped electrons originate from the innermost nitrogen states, which are only ionized near the peak of the laser pulse intensity. These electrons then drift to the back of the bubble and are accelerated forward by the joint LWFA and DLA processes. All PIC simulations presented in this work are performed using \textsc{Osiris}.\cite{Fonseca2002,Davidson2015}

\subsection{Electron beam characteristics and impact of field solver} \label{sec:pic-integrated}

\begin{table}[]
\begin{tabular}{l | l }
\hline \hline
Laser wavelength ($\lambda$)     & 815$\,$nm  \\
\hline
Pulse duration (FWHM) & 45$\,$fs \\
\hline
Spot size ($w_0$) & 6.7$\,\mu$m \\
\hline
Amplitude ($a_0$) & 2.03 \\
\hline \hline
Electron density ($n_0$) & $1.43\times 10^{19}\,$cm$^{-3}$    \\
\hline
Acceleration distance & 430$\,\mu$m          \\
\hline \hline                                    
\end{tabular}
\caption{Physical parameters used for the PIC simulations. The simulations are based on an experiment described in Ref.~\onlinecite{Shaw2017}, in which significant DLA was observed in an LWFA.}
\label{tab:params}
\end{table}

\begin{figure}
\includegraphics[width=\linewidth]{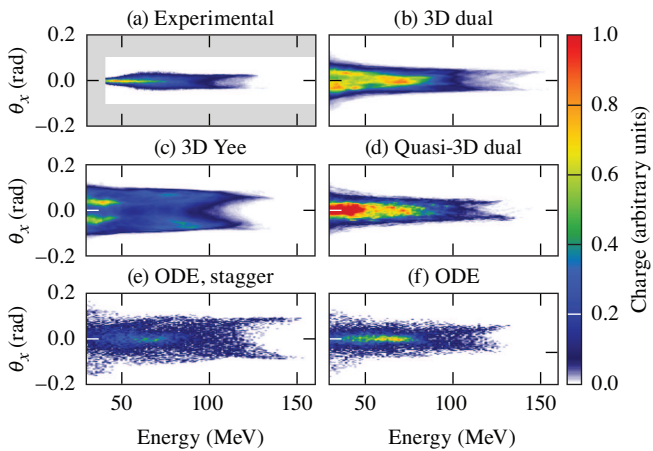}
\caption{\label{fig:gamma-theta} The final electron energy and divergence angle from (a)~the experimental data presented in Ref.~\onlinecite{Shaw2017}; PIC simulations using (b)~the 3D dual solver, (c)~the 3D Yee solver, and (d)~the dual solver in quasi-3D; and the numerically integrated equations of motion (e)~with and (f)~without the time-staggering error. The simulations using the dual solver are in much better agreement with the experiment than those using the Yee solver.}
\end{figure}

Figure~\ref{fig:gamma-theta} compares the experimentally measured and predicted energy-dependent divergence angles, $\theta_x = \mathrm{atan}(p_x/p_z)$, of accelerated electrons. The simulations using the dual solver are in much better agreement with the experiment than the simulation using the Yee solver. While each case exhibits the ``forking'' characteristic of DLA,\cite{Shaw2017,Shaw2018,Gong2020DirectFields,King2021} i.e., the absence of charge at small angles for energies greater than ${\sim}90$~MeV, using the Yee solver results in far fewer electrons with small divergence angles, especially at lower energies. For this comparison, the PIC simulations use 30 points per $\lambda = 2\pi c/\omega_0$ and a time step set close to the CFL limit for each solver.

To validate and further explore the observed differences in the PIC simulations, Eq.~(\ref{eq:motionchannelreduced}), which models electron motion in the combined laser pulse and bubble fields, is numerically integrated. The effect of time staggering that appears in Eq.~(\ref{eq:Lorentz-fouier}) is included in Eq.~(\ref{eq:motionchannelreduced}) by multiplying the magnetic field terms, i.e., the terms proportional to $\bold{v}\sin(k_z \xi)$, by $\cos (\omega_0 \Delta t/2)$. Despite being close to unity [$\cos (\omega_0 \Delta t/2)\approx 0.994$ for the Yee-solver case], this factor results in a substantial increase in the divergence angle of the highest-energy electrons [cf. Figs.~\ref{fig:gamma-theta}(e) and \ref{fig:gamma-theta}(f)]. Similar behavior is observed in the PIC simulation employing the Yee solver [Fig.~\ref{fig:gamma-theta}(c)], suggesting that the time-staggering error artificially amplifies the transverse oscillations of electrons in the fields of the laser pulse.

\begin{figure}
    \centering
    \includegraphics[width=\linewidth]{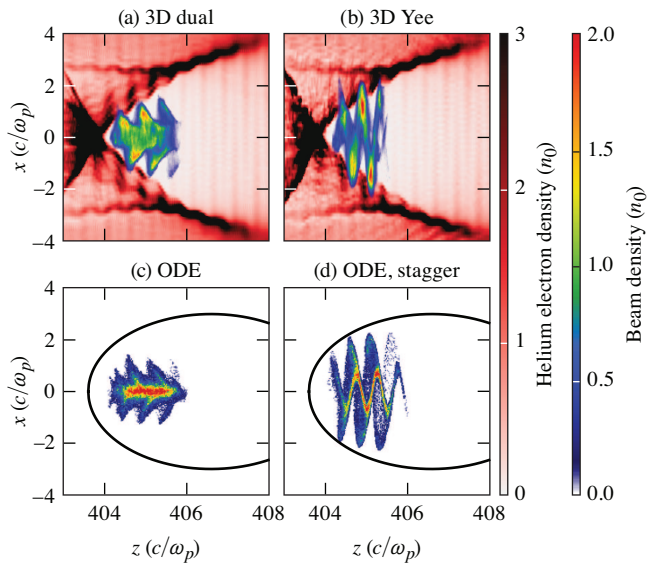}
    \caption{\label{fig:beam-position} The electron beam density predicted by 3D PIC simulations using (a)~the dual solver and (b)~the Yee solver and the numerically integrated equations of motion (c)~with and (d)~without the time-staggering error. The profiles are shown near the end of the acceleration, just before the density down ramp. The density of electrons originating from helium appears in the background of (a) and (b), illustrating the structure of the bubble. The time-staggering error produces unphysically large transverse oscillations that extend well beyond the bubble.}
\end{figure}

Figure~\ref{fig:beam-position} displays the spatial distribution of the ionization-injected electrons predicted by the 3D PIC simulations and by Eq.~(\ref{eq:motionchannelreduced}). In all cases, the electron density exhibits a sinusoidal modulation at half the laser wavelength that results from resonantly driven betatron oscillations. Consistent with the artificially enhanced angles observed in Fig.~\ref{fig:gamma-theta}(c), the Yee-solver simulation in Fig.~\ref{fig:beam-position}(b) produces transverse oscillations with an unphysically large amplitude, extending well outside of the bubble (visible as the density contours of electrons originating from He). Including the time-staggering factor in Eq.~(\ref{eq:motionchannelreduced}) produces similar behavior [Fig.~\ref{fig:beam-position}(d)]. In contrast, the PIC simulation using the dual solver and Eq.~(\ref{eq:motionchannelreduced}) without the time-staggering coefficient produces a wedge-shaped structure with more of the electrons found near the axis. 

\subsection{Convergence of the Yee and dual solvers}

\begin{figure}
    \centering
    \includegraphics[width=\linewidth]{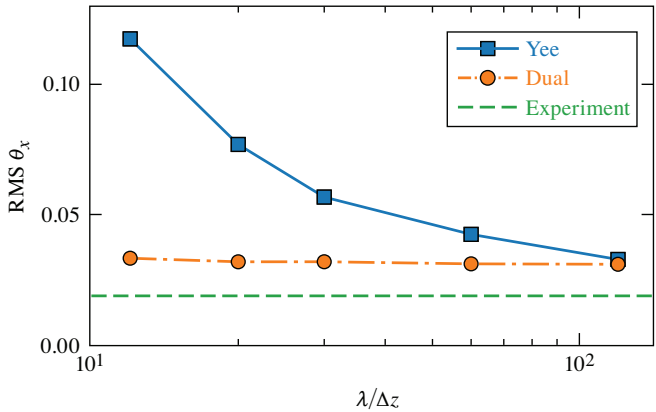}
    \caption{\label{fig:convergence} The root-mean-square divergence angle of the highest-energy electrons from 3D PIC simulations using the Yee and dual field solvers as a function of longitudinal resolution, together with the experimental data. The dual solver provides convergence at a much lower resolution than the Yee solver. }
\end{figure}

In order to eliminate the discrepancies between the simulations using the dual solver and Yee solver, the spatial resolution used with the Yee solver must be increased by an order of magnitude. Figure~\ref{fig:convergence} displays the results of a convergence test comparing the root-mean-squared (RMS) divergence angle predicted by the dual- and Yee-solver simulations. The RMS angle is nearly identical for all simulations using the dual solver, whereas converged simulations using the Yee solver require a much smaller longitudinal cell size. In all cases, the simulations are run with a time step close to the CFL limit, and the number of particles per cell is kept constant at 8 (except the largest run with 120 points per wavelength, where 4 particles per cell are used). For the example considered, a spatial resolution of 30 points per $\lambda$ is sufficient for convergence of the dual-solver simulations but is insufficient for the Yee solver. The remaining discrepancy between the converged PIC and experimental results can be attributed to differences in the profile of the laser pulse and gas or plasma conditions.

The errors in the dispersion relation and Lorentz force inherent to the Yee solver are quadratic in $\Delta t$,\cite{Li2021} meaning that convergence is expected as the time step is reduced. However, to ensure physically accurate dispersion, as determined by the CFL condition, the spatial step must be reduced as well. This is in contrast to the dual solver, where the coefficients of the spatial derivatives can be customized to ensure accurate dispersion for any time step, which allows for convergence by reducing the time step alone. The tradeoff is that the dual solver requires a larger stencil for the spatial derivatives and has a stricter CFL condition (about 70\% that of the Yee solver for the same grid size).

In terms of overall simulation performance, the additional operations introduced by the larger stencil of the dual solver are more than offset by the reduction in spatial grid points and time steps needed for convergence. By requiring fewer spatial grid points, the dual solver cuts down on the memory footprint and total number of particles (and particle operations). As a result, the dual solver ultimately provides a savings in total computation time. In the specific cases presented here, the converged dual solver simulations (30 points per $\lambda$) take ${\sim}6\times$ less time to complete than the converged Yee-solver simulations (120 points per $\lambda$). For comparison, when the same longitudinal resolution is used, the dual-solver simulations take 1.5 to 2.5$\times$ longer than the Yee-solver simulations. 

Notably, Fig.~\ref{fig:convergence} suggests that many of the state-of-the-art simulations that employ traditional solvers\cite{Nemeth2008,Martins2010,Yin2012TrappingBeams,Shaw2017,Snyder2019RelativisticTargets,Yoffe2020Particle-in-cellWindow} may be using insufficient resolution for convergence. While the dual solver can improve the accuracy and performance of any PIC simulation of laser--particle interactions,\cite{Li2021} the benefits are most prominent when the particles spend significant time within the fields of the laser pulse as in LWFA and SM-LWFA.

\subsection{Energy gain mechanisms and azimuthal-mode decomposition} \label{sec:pic-energy}
Quantifying the contributions of different acceleration mechanisms to the electron energy gain allows for more-informed designs and optimization of an LWFA. The quasi-3D geometry enables unambiguous calculation of these contributions. More specifically, the expansion of the electromagnetic fields into azimuthal modes completely separates the fields of the plasma wave and the laser pulse. 

As discussed in Sec.~\ref{sec:quasi-3D}, the energy gain contributions from DLA and LWFA are often calculated by attributing the work done by the longitudinal fields to LWFA and the work done by the transverse fields to DLA. However, this method can be inaccurate because finite-sized laser pulses have a longitudinal electric field, which, in the case of linear polarization, always decelerates electrons in betatron resonance.\cite{Quesnel1998,Pukhov2003} Thus, using only the transverse fields to approximate the work done by DLA can overestimate the DLA contribution to the energy gain. A more-accurate method corrects the work done by the fields with an approximation for the energy loss due to the longitudinal field of the laser pulse as in Eqs.~(\ref{eq:f}) and (\ref{eq:work-f}). Nevertheless, neither of these methods are as accurate as using the quasi-3D decomposition. 

\begin{figure}
    \centering
    \includegraphics[width=\linewidth]{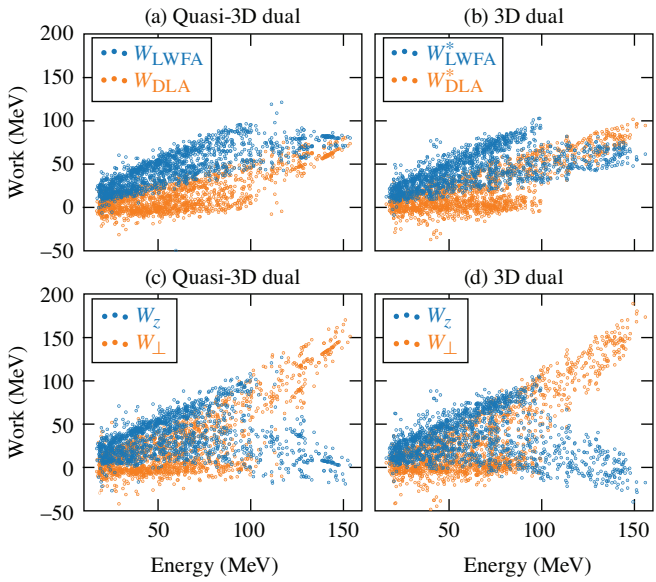}
    \caption{\label{fig:modal-transv-work} A comparison of the work done by DLA and LWFA calculated using different methods: (a)~quasi-3D decomposition [Eq.~(\ref{eq:work-modal})], (b)~3D using the longitudinal and transverse fields with an approximate correction for the longitudinal field of the laser pulse [Eq.~(\ref{eq:work-f})], and using the longitudinal and transverse fields without the correction [Eq.~(\ref{eq:work-trad})] for (c)~quasi-3D and (d)~3D simulations. The dual solver is used for each case, and $1500$ particles are tracked to calculate the work. The agreement between (c) and (d) gives confidence that the quasi-3D simulations produce the correct physics. The quasi-3D decomposition in (a) is most accurate and reveals that the highest-energy electrons receive comparable energy from LWFA and DLA.}
\end{figure}

Figure~\ref{fig:modal-transv-work} illustrates how each method gives a different prediction for the work done by LWFA and DLA. The most accurate method, based on the quasi-3D decomposition [Eq.~(\ref{eq:work-modal})], demonstrates that the highest-energy electrons receive comparable energy from LWFA and DLA [Fig.~\ref{fig:modal-transv-work}(a)]. The least accurate method, which uses the transverse and longitudinal fields [Eq.~(\ref{eq:work-trad})], significantly over and underestimates the DLA and LWFA contributions, respectively [Figs.~\ref{fig:modal-transv-work}(c) and ~\ref{fig:modal-transv-work}(d)]. Applying a correction to this method that accounts for the work done by the longitudinal field of the laser pulse provides much better agreement with the quasi-3D result, but still over and underestimates the DLA and LFWA contributions [Fig.~\ref{fig:modal-transv-work}(b)]. The work done by the longitudinal field of the pulse is calculated using $\omega_0=12\,\omega_p$ and $w_0=2.53\,c/\omega_p$ in Eqs.~(\ref{eq:f}) and (\ref{eq:work-f}), which are average values near the electron bunch. Generally, these quantities can be dynamic in nature and difficult to estimate, which makes the correction method relatively unreliable.

\begin{figure}
    \centering
    \includegraphics[width=\linewidth]{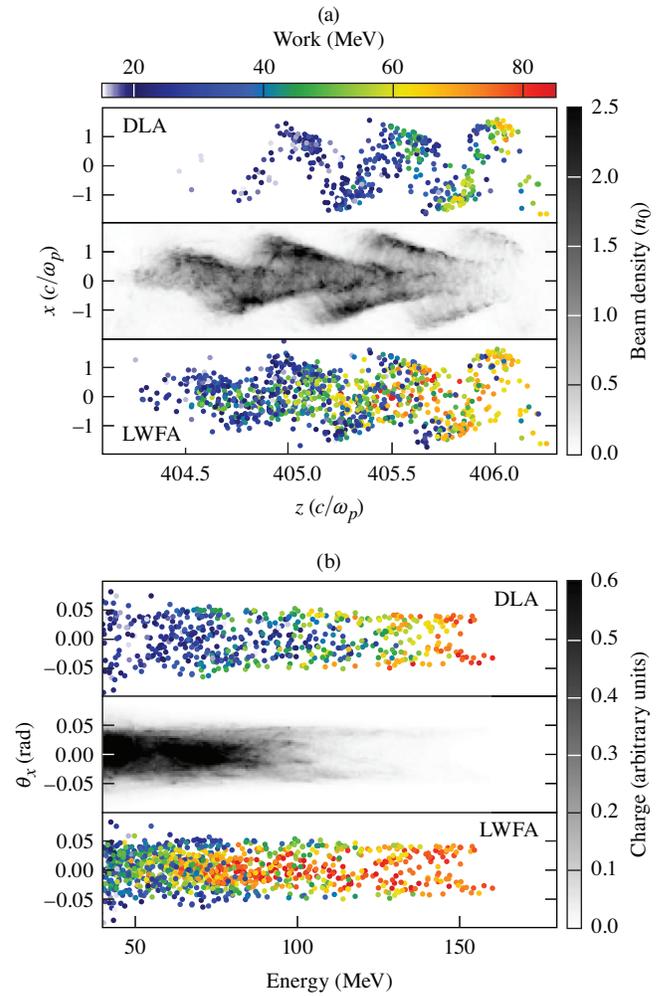}
    \caption{\label{fig:sim-work} (a)~Density of the accelerated electron bunch near the end of the plasma, just before the background density down ramp.  (b)~The final energy-dependent divergence angle of the electron bunch. The top and bottom panels in (a) and (b) show the work done by DLA and LWFA, respectively [Eq.~(\ref{eq:work-modal})]. The data is from the quasi-3D PIC simulation using the dual solver. The highest-energy electrons, which receive comparable energy from LWFA and DLA, tend to have larger divergence angles due to their large-amplitude betatron oscillations.}
\end{figure}

The accurate partitioning of work afforded by the quasi-3D method can reveal additional insights into the DLA and LWFA processes. Figure~\ref{fig:sim-work} displays the work done by DLA (top panels) and LWFA (bottom panels) overlayed on (a)~the beam density near the end of acceleration and (b)~the energy-dependent divergence angle at the end of the simulation. As observed in Fig.~\ref{fig:sim-work}(a), LWFA accelerates all of the high-energy electrons to some degree. In contrast, DLA only occurs for electrons executing betatron motion with large transverse excursions from the propagation axis. In both cases, the energy gain increases with longitudinal position, correlating with time spent in the wake. Consistent with the large transverse excursions, electrons that receive comparable energy from DLA and LWFA generally have larger divergence angles [Fig.~\ref{fig:sim-work}(b)], while those that gain most of their energy from LWFA have smaller divergence angles.

\section{Conclusion} \label{sec:conclusion}
Two techniques have been presented that improve the predictions of PIC simulations when modeling DLA-assisted LWFA. The customized (``dual'') solver\cite{Li2021} corrects for errors in the dispersion relation of light waves and in the Lorentz force due to the time staggering of electric and magnetic fields. The quasi-3D algorithm enables unambiguous analysis of the energy gain contributions from DLA and LWFA. Together, these methods  expedite analysis, optimization, and design of LWFA experiments. 

The benefits of these techniques were illustrated by simulations of LWFA experiments in which DLA contributed significantly to the energy gain of ionization-injected electrons. For typical resolutions, the results of simulations employing the dual solver were in much better agreement with the experimental data than those employing the Yee solver. The Yee-solver simulations predicted artificially large transverse momenta for electrons in betatron resonance. This is a direct result of the time-staggering error, which was demonstrated using the numerically integrated equations of motion with a factor that accounted for this error. With an order-of-magnitude increase in the spatial resolution, the Yee-solver simulations eventually converged, but at the cost of an ${\sim}6\times$ increase in the run time compared to the converged dual-solver simulations.

The quasi-3D decomposition revealed two distinct populations of high-energy electrons. One population gained nearly all of its energy from the wakefield, remained localized near the bubble axis, and had small divergence angles. The other population received comparable energy from the DLA and LWFA processes, was distributed in a wedge-like, sinusoidal pattern in the bubble, and had larger divergence angles. This observation is in stark contrast to the conclusion one would draw by attributing the energy gain from DLA and LWFA to the transverse and longitudinal fields, respectively. Specifically, this method would incorrectly predict that the highest-energy electrons receive most of their energy from DLA.

Either individually or combined, the dual-solver and quasi-3D geometry can facilitate further investigations of LFWA, DLA-assisted LWFA, and SM-LWFA. Examples include enhancing DLA or sustaining betatron resonance via tailored density or laser pulse profiles, accelerating electrons to higher harmonics of the betatron resonance, increasing the emitted betatron radiation, or decreasing the beam emittance. More generally, the dual solver may increase performance in any system where the particles interact with a laser pulse over long durations and distances. In these cases, the quasi-3D algorithm may provide the insight needed to fully understand the processes contributing to energy gain.

\begin{acknowledgments}
This report was prepared as an account of work sponsored by an agency of the U.S. Government. Neither the U.S. Government nor any agency thereof, nor any of their employees, makes any warranty, express or implied, or assumes any legal liability or responsibility for the accuracy, completeness, or usefulness of any information, apparatus, product, or process disclosed, or represents that its use would not infringe privately owned rights. Reference herein to any specific commercial product, process, or service by trade name, trademark, manufacturer, or otherwise does not necessarily constitute or imply its endorsement, recommendation, or favoring by the U.S. Government or any agency thereof. The views and opinions of authors expressed herein do not necessarily state or reflect those of the U.S. Government or any agency thereof.

This material is based upon work supported by the Office of Fusion Energy Sciences under Award Number DE-SC00215057, the Department of Energy National Nuclear Security Administration under Award Number DE-NA0003856, the University of Rochester, and the New York State Energy Research and Development Authority. Additional support was given by DOE grant DE-SC0019010 and NSF grant 1806046.  Simulations were performed at NERSC under m1157 and m3013.
\end{acknowledgments}

\section*{Data Availability Statement}

The data that support the findings of this study are available from the corresponding author upon reasonable request.

\appendix

\section{Simulation parameters} \label{app:params}
In this appendix, details are given for the PIC and numerical integration simulations discussed in Sec.~\ref{sec:results}.  The plasma density in the PIC simulation comprises a 100-$\mu$m density up ramp (the laser is focused halfway through the up ramp), a 430-$\mu$m constant-density region, and a 150-$\mu$m density down ramp.  The grid is $1814\times320\times320$ with eight particles per cell for each species ($1814\times160$ and 32~particles per cell per species in quasi-3D) with 30~points per laser wavelength in the longitudinal direction and 52~points per plasma period in the transverse directions.  The time step for each simulation is 0.01875$\,\omega_p^{-1}$, 0.01295$\,\omega_p^{-1}$ and 0.0111$\,\omega_p^{-1}$ for the 3D Yee, 3D dual, and quasi-3D dual simulations, respectively. Simulations employing the dual solver use 16 coefficients for the finite-difference operators (the standard Yee solver uses one coefficient).

The numerical integration of the ordinary differential equations in Eq.~(\ref{eq:motionchannelreduced}) is performed with an explicit fifth-order Runge--Kutta method.  Using the observed injection time from the PIC simulation as a guide, particles are injected into the bubble over a time of 200$\,\omega_p^{-1}$.  These particles are injected on-axis, with $\zeta$ chosen randomly from the interval $\zeta \in [-2R+0.286\,c/\omega_p,-2R+0.836\,c/\omega_p]$ and $x$ from the interval $x \in [-r_b(\zeta),r_b(\zeta)]$, where $r_b(\zeta) = [R^2-(\zeta+R)^2]^{1/2}$ is the assumed bubble radius as a function of $\zeta$.  The initial particle energy is randomly selected from $\gamma \in [\gamma_g,2\gamma_g]$, where $\gamma_g = [1-(\mathrm{v}_g/c)^2]^{-1/2}$.  The angle that the initial velocity makes with respect to the $x$~axis is randomly chosen with the constraint that $\mathrm{v}_z \geq \mathrm{v}_g$.  All cases use $\mathrm{v}_\phi = 1.002565\,c$, $\mathrm{v}_g = 0.993583\,c$, $k_z = 12.14\, \omega_p/c$, $R=3\,c/\omega_p$ and $a_0 = 2.8$, which are measured from the 3D PIC simulations.  Note that the laser wavenumber and $a_0$ are larger than those initialized in the PIC simulation due to photon acceleration and self-focusing, respectively.

\bibliography{references}

\end{document}